\documentclass{aa}
\usepackage{graphicx}
\usepackage{txfonts}
\newcommand{\Si}[5]{\mbox{$#1\,^#2{\rm #3}^{{\rm #4}}_{\rm #5}$}}

\newcommand{\SH}{$S_{\!\!\rm H}$}

\begin{document}
\title{\textbf{Statistical equilibrium of silicon in the atmospheres of metal-poor stars}
\thanks{Based on observations obtained in the frame of the ESO programme
ID 165.N-0276(A).}}
%\thanks{Tables 2 are only available in electronic form at
%http://www.edpsciences.org}
\author{J.R. Shi\inst{1,2} \and T. Gehren\inst{2} \and L. Mashonkina\inst{2,3} \and G. Zhao\inst{1}} \offprints{J.R. Shi,\ \
\email{sjr@bao.ac.cn}}

\institute {National Astronomical Observatories, Chinese Academy of
Sciences, Beijing 100012, P. R. China \and Universit\"ats-Sternwarte
M\"unchen, Scheinerstrasse 1, D-81679 M\"unchen, Germany \and
Institute of Astronomy, Russian Academy of Sciences, Pyatnitskaya
Str. 48, 109017 Moscow, Russia\\}
\date{Received / Accepted }

\abstract
% context heading (optional)
{}
% aims heading (mandatory)
{The statistical equilibrium of neutral and ionized silicon in the
atmospheres of metal-poor stars is discussed. Non-local
thermodynamic equilibrium effects are investigated and the silicon
abundances in metal-poor stars determined.}
% methods heading (mandatory)
{We have used high resolution, high signal to noise ratio spectra
from the UVES spectragraph at the ESO VLT telescope. Line formation
calculations of \ion{Si}{i} and \ion{Si}{ii} in the atmospheres of
metal-poor stars are presented for atomic models of silicon
including 174 terms and 1132 line transitions. Recent improved
calculations of \ion{Si}{i} and \ion{Si}{ii} photoionization
cross-sections are taken into account, and the influence of the
free-free quasi-molecular absorption in the Ly$\alpha$ wing is
investigated by comparing theoretical and observed fluxes of
metal-poor stars. All abundance results are derived from LTE and
NLTE statistical equilibrium calculations and spectrum synthesis
methods.}
% results heading (mandatory)
{It is found that the extreme ultraviolet radiation is very
important for metal-poor stars, especially for the high temperature,
very metal-poor stars. The radiative bound-free cross-sections also
play a very important role for these stars.}
% conclusions heading (optional)
{NLTE effects for Si are found to be important for metal-poor stars,
in particular for warm metal-poor stars. It is found that these
effects depend on the temperature. For warm metal-poor stars, the
NLTE abundance correction reaches $\sim$ 0.2 dex relative to
standard LTE calculations. Our results indicate that Si is
overabundant for metal-poor stars.} {}

 \maketitle

\section{Introduction}

Silicon is not only an important reference element for comparision
of various types of cosmic matter with the Sun, but also one of the
main electron contributors and opacity sources in the near UV in the
atmospheres of cool stars. With few exceptions, silicon also
represents the only $\alpha$-element measurement for a given damped
Ly$\alpha$ system (DLA, Prochaska \& Wolfe \cite{PW02}). Although
silicon is a refractory element, its depletion is mild in lightly
depleted regions of the ISM, and therefore the observed column
densities for the DLA system should nearly reflect the total Si
column density.

The current paradigm is that silicon is made during oxygen and neon
burning in massive stars, and it would therefore be ejected by Type
II supernove (Woosley \& Weaver \cite{WW95}; Umeda et al.
\cite{UNN00}; Ohkubo et al. \cite{O06}). Type Ia supernove also
produce some silicon (Tsujimoto et al. \cite{TNY95}; Iwamoto et al.
\cite{IBN99}).
%\textbf{In the latter case $^{23}$Na
%%metal-poor stars.}

\begin{figure}
\resizebox{\hsize}{7.0cm}{\includegraphics[width=9.3cm]{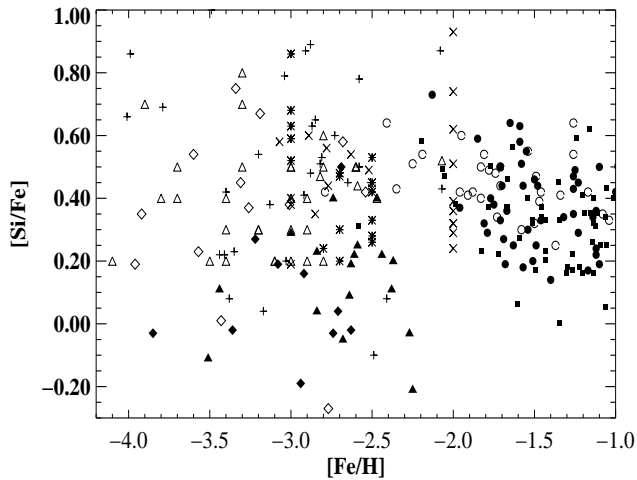}}
\caption[short title]{Silicon abundances of metal-poor stars
determined from LTE analyses of Fulbright (\cite{Fu00}, Filled
circles $\bullet$ for dwarfs and open circles $\circ$ for giants),
McWilliam et al. (\cite{MPS95}, pluses $+$), Honda et al.
(\cite{H04}, crosses $\times$), Aoki et al. (\cite{A05}, asterisks
$\ast$), Ryan et al. (\cite{RNB96}, filled diamonds
$\blacklozenge$ for dwarfs and open diamonds $\diamond$ for
giants), Gratton et al. (\cite{GCC03}, filled squares
$\blacksquare$), Cayrel et al. (\cite{CDS04}, open triangles
$\triangle$), and Cohen et al. (\cite{CCM04}, filled triangles
$\blacktriangle$).} \vspace*{-0.3cm} \label{fig1}
\end{figure}

Previous abundance determinations of Si have been carried out under
the assumption of local thermodynamic equilibrium (LTE). They show
pronounced star-to-star scatter at all metallicities, and
particularly for extremely metal-poor stars, where the observed Si
abundance is represented by only two lines and affected by the
contamination of CH and H$_\delta$ lines. This is seen in
Fig.\ref{fig1}. Preston et al. (\cite{PST06}) and Lai et al.
(\cite{LBJ08}) found that the derived Si abundance depends only on
$T_{\rm eff}$ and not on log $g$, and Preston et al. (\cite{PST06})
suggested that the Si abundances for stars with $T_{\rm eff}$ $>$
5800 K probably do not represent the true values. Cohen et al.
(\cite{CCM04}) noted that the observational picture for Si shows a
puzzle -- the abundance ratio [Si/Fe] of giants is about 0.4\,dex
larger than that obtained for dwarfs (also see Ryan et al.
\cite{RNB96}), and they discussed in detail the possibility of this
difference:
\begin{itemize}
\item There is a problem for the transition probabilities of the
two very strong \ion{Si}{i} lines. Cohen et al. (\cite{CCM04})
studied high temperature extremely metal-poor dwarfs, but they
used only the 3905 \AA\ \ion{Si}{i} line. This line overlaps a CH
feature; however, it is very weak in such stars. Cayrel et al.
(\cite{CDS04}) relied instead on a single line of \ion{Si}{i} at
4102.9 \AA, which is close to H$_\delta$. In cooler giants the
3905 \AA\ line is too blended with CH even in normal carbon
abundance stars. Both Cohen et al. (\cite{CCM04}) and Cayrel et
al. (\cite{CDS04}) adopted the log $gf$ value from the laboratory
work of Garz (\cite{Garz73}). However, NIST gives -2.92\,dex for
the 4102 \AA\ line, which is 0.22 \,dex higher than the laboratory
value. Thus, the Si abundance derived by Cayrel et al.
(\cite{CDS04}) would be 0.22 dex lower if the NIST $gf$ value were
adopted.
%Cohen et al. (\cite{CCM04})for
%this line, than that of
% for the same line strength.

\item
%There is the issue of NLTE.
Only one non-local thermodynamic equilibrium (NLTE) analysis has
been published to date, in which Wedemeyer (\cite{WE01}) reported
the influence of deviations from LTE upon the Si abundances in the
Sun and Vega, with no reference to line formation in metal-poor
stars; he demonstrated that NLTE for Si in the Sun is negligibly
small. However, he did not consider line formation for the two
strong lines.
\end{itemize}
Cohen et al. (\cite{CCM04}) suggested that this difference probably
arises from contamination of the \ion{Si}{i} line at 3905 \AA\ by
blending CH features in the giants; spectral syntheses were not
carried out for the giants by Cayrel et al. (\cite{CDS04}). Preston
et al. (\cite{PST06}) examined this suggestion in their red
horizontal-branch (RHB) stars and found that the CH band is very
weak. Testing synthetic spectra of cooler RHB stars computed using
derived carbon abundances revealed that the CH contribution can only
be a few percent of the total absorption at the \ion{Si}{i} feature.
They also found that the derived Si abundances for cooler stars are
in reasonable accord with values determined for stars of similar
metallicity using \ion{Si}{i} transitions in the red spectral
region.

The present work is based on a sample of metal-poor stars and aims
at exploring their [Si/Fe] abundance ratios applying a full spectrum
synthesis based on level populations calculated from the statistical
equilibrium equations. In Sect. 2 we present the observational
technique and the atmospheric models and stellar parameters are
discussed in Sect. 3. NLTE line formation is discussed in Sect. 4.
The results and comparison with other works are shown in Sec. 5. The
discussion is presented in Sect. 6, and the conclusions are given in
Sect. 7.

\section{Observations}

Our approach to obtain a representative silicon abundance
investigation involves analyzing a sample of metal-poor stars with
high resolution and high signal to noise ratio spectra. The spectra
of 13 metal-poor dwarf stars were observed in March 2001 with the
UVES \'{e}chelle spectrograph mounted at the ESO VLT2. UVES
observations cover a spectral range between 3300 and 6650 \AA\ with
gaps of $\sim$100 \AA\ around 4570 \AA\ due to the beam splitter and
near 5580 \AA\ at the edge of the butted CCD. The spectral
resolution is around R = 60\,000. These observations were originally
intended to show a high signal in the blue (see Mashonkina et al.
2003). Consequently, the green/red spectra have a S/N near 300 in
most of the single exposures. We also use high-quality observed
spectra from the ESO UVESPOP survey (Bagnulo et al. \cite{BJL05})
for HD\,122563. For G\,64-12 we use the spectrum from the High
Dispersion Spectrograph at the Nasmyth focus of the Subaru 8.2 m
telescope (Noguchi et al. \cite{NAK02}).

Data extraction followed the standard automatic IDL program
environment designed for the FOCES spectrograph (Pfeiffer et al.
1998), but with slight modifications also applicable to the UVES
data. All \'{e}chelle images including flat field and ThAr were
corrected for bias and scattered light background. Objects and ThAr
exposures were extracted and corrected for flat field response. Bad
pixels were detected and as far as possible removed by comparison of
the 3 single exposures (see Gehren et al. \cite{GLS04} for detail).

%\section {Atmospheric models and stellar parameters}
\section{Method of calculation}

\subsection{Model atmospheres}

Our analyses are all based on the same type of atmospheric model,
irrespective of temperature, gravity or metal abundance. We use
line-blanketed LTE model atmospheres, generated as discussed by
Fuhrmann et al. (\cite{FP97}). The main characteristics are: the
iron opacity was calculated with the improved meteorite value log
$\varepsilon_{\rm Fe} = 7.51$ (Anders \& Grevesse 1989); opacities
for metal-poor stars with [Fe/H] $< -0.6$ were calculated using
$\alpha$-element (O, Mg, Si and Ca) abundances enhanced by 0.4 dex,
and the mixing-length parameter $l$/H$_p$ was adopted to be 0.5, in
order to determine consistent temperatures for H$_\alpha$ and the
higher Balmer lines (see Fuhrmann et al. \cite{FAG93}).

\subsection{Stellar parameters}

For most of our program stars, we adopted the stellar parameters
determined by Gehren et al. (\cite{GLS04}, \cite{GSZ06}), where the
effective temperatures are derived from the wings of the Balmer
lines; surface gravities are based on the HIPPARCOS parallaxes. Iron
abundances are obtained from \ion{Fe}{ii} lines, and the
microturbulence velocities are estimated by requiring that the iron
abundance derived from \ion{Fe}{ii} lines should not depend on
equivalent width. For HD\,84937 and HD\,122563 the parameters were
taken from Mashonkina et al. (\cite{MZG08}), where the stellar
effective temperatures were determined from the hydrogen H$_\alpha$
and H$_\beta$ line wing fitting based on NLTE line formation;
surface gravities are based on the HIPPARCOS parallaxes, and iron
abundances are derived from \ion{Fe}{ii} lines and the
microturbulence velocity was derived from the strongest
\ion{Fe}{ii}, \ion{Ca}{i}, and \ion{Mg}{i} lines. For consistency
with the whole series of our earlier NLTE studies (from Baum\"uller
\& Gehren \cite{BG96} to the most recent paper of Shi et al.
\cite{SGB08}), we adopt effective temperatures based on the hydrogen
resonance broadening calculated with Ali \& Griem (1966) theory. We
note that two papers of Barklem et al. \cite{BPO00} and Allard et
al. (\cite{A08}) obtain larger values of the hydrogen
self-broadening cross-sections. An impact of using the
self-broadening formalism of Barklem et al. (\cite{BPO00}) on
effective temperature determinations for metal-poor stars is
discussed by Mashonkina et al. (\cite{MZG08}). The uncertainties for
the temperature, surface gravity, metal abundance and
microturbulence velocities are generally assumed to be $\pm$50 K,
0.05 dex, 0.05 dex and 0.1 km s$^{-1}$ respectively.

\subsection {Atomic line data}

Table \ref{table1} lists the relevant line data with their final
solar fit values (Shi et al. \cite{SGB08}, hereafter paper I).
Collisional broadening through van der Waals interaction with
hydrogen atoms is important for strong Si lines. As already pointed
out by Gehren et al. (\cite{GBM01}, \cite{GLS04}), the resulting
values of the van der Waals damping constants are mostly near those
calculated according to Anstee \& O'Mara's (\cite{AO91},
\cite{AO95}) tables. In our analysis, the absolute value of the
oscillator strengths is unimportant because the abundances are
evaluated in a fully differential way with respect to the sun.

\begin{table}
\caption[1]{Atomic data of silicon lines$^*$.}
%\centering{
\begin{tabular}{rr@{ $-$ }lrr}
\hline\hline\noalign{\smallskip}
$\lambda$~ [\AA] & \multicolumn{2}{c}{Transition} & $\log gf$ & $\log C_6$\\
\hline\noalign{\smallskip}
 3905.53 &  \Si{3p}{1}{S}{}{0}   & \Si{4s}{1}{P}{o}{1} & -1.10  & -30.917  \\
 4102.93 &  \Si{3p}{1}{S}{}{0}   & \Si{4s}{3}{P}{o}{1} & -2.99  & -30.972  \\
 5690.43 &  \Si{4s}{3}{P}{o}{1}  & \Si{5p}{3}{P}{}{1}  & -1.74  & -30.294  \\
 5701.11 &  \Si{4s}{3}{P}{o}{1}  & \Si{5p}{3}{P}{}{0}  & -1.96  & -30.294  \\
 5772.15 &  \Si{4s}{1}{P}{o}{1}  & \Si{5p}{1}{S}{}{0}  & -1.62  & -30.287  \\
 6142.49 &  \Si{3p^3}{3}{D}{o}{3}& \Si{5f}{3}{D}{}{3}  & -1.48  & -29.869  \\
 6145.05 &  \Si{3p^3}{3}{D}{o}{2}& \Si{5f}{3}{G}{}{3}  & -1.39  & -29.869  \\
 6155.14 &  \Si{3p^3}{3}{D}{o}{3}& \Si{5f}{3}{G}{}{4}  & -0.78  & -29.869  \\
 6237.32 &  \Si{3p^3}{3}{D}{o}{1}& \Si{5f}{3}{F}{}{2}  & -1.08  & -29.869  \\
 6243.82 &  \Si{3p^3}{3}{D}{o}{2}& \Si{5f}{3}{F}{}{3}  & -1.29  & -29.869  \\
 6244.47 &  \Si{3p^3}{3}{D}{o}{2}& \Si{5f}{1}{D}{}{2}  & -1.29  & -29.869  \\
\noalign{\smallskip}\hline\noalign{\smallskip}
 6347.10 &  \Si{4s}{2}{S}{}{1/2} & \Si{4p}{2}{P}{o}{3/2}&  0.26  & -30.200  \\
 6371.36 &  \Si{4s}{2}{S}{}{1/2} & \Si{4p}{2}{P}{o}{1/2}& -0.06  & -30.200  \\
\noalign{\smallskip} \hline
\end{tabular}
%}

$^*$$\log gf$ values have been determined from solar spectrum fits,
and damping constants log $C_6$ for \ion{Si}{i} lines are computed
according to the Anstee \& O`Mara (1991, 1995) interpolation tables.
\label{table1}
\end{table}

\section{NLTE calculations}

\subsection{Atomic model}

The silicon model atom includes the most important levels of \ion
{Si}{i} and \ion {Si}{ii} and comprises 132 terms of \ion {Si}{i},
41 terms of \ion{Si}{ii}, plus the \ion{Si}{iii} ground state. The
atomic properties are documented in Shi et al. (paper I). Oscillator
strengths and photoionization cross-sections were calculated by
Nahar \& Pradhan (\cite{NP93}). The full analysis of the solar
spectrum (Kurucz et al. \cite{KFB84}) allows a reasonable choice of
the hydrogen collision enhancement factor resulting in \SH\ = 0.1
for \ion{Si} (see paper I).

All calculations are carried out with a revised version of the
DETAIL program (Butler \& Giddings \cite{BG85}) using accelerated
lambda iteration (see Gehren et al. {\cite{GBM01}, \cite{GLS04} for
details).

\subsection{NLTE effects}

The abundance analyses of silicon clearly show the NLTE effect.
There is a tendency that the NLTE effect is large for warm
metal-poor stars, as would be expected.
%Baum\"{u}ller et al.(\cite{BBG98})
Our results confirm that LTE abundances can be significantly
different from their NLTE counterparts, with differences reaching
more than 0.20 dex in extreme cases. It is evident that the NLTE
effects are systematically stronger for hotter models, which is in
agreement with the statistical equilibrium of sodium, magnesium
and aluminium (Shi et al. \cite{SGZ04}; Zhao \& Gehren
\cite{ZG00}; Baum\"{u}ller et al. \cite{BG97}). As expected, the
strongest departures from LTE are found for models with high
temperature and low metallicity. For our program stars, the
differences of abundance between LTE and NLTE analyses for the
strong 3905 and 4102 \AA\ lines are plotted in Fig. \ref{fig2} as
a function of metal abundance, temperature and surface gravity,
respectively.

It should be noted that the NLTE effects differ from line to line,
reflecting their individual properties. For example, the strong 3905
and 4102 \AA\ lines show large NLTE effects compared to the other
weak lines. From Table \ref{table2}, we can see the importance of
the NLTE effects on different lines in the silicon abundance
determination:

\begin{itemize}
\item the weak lines show the smallest NLTE abundance effects ($< 0.02$ dex) for
the stars studied here, so they are the best abundance indicators
for LTE analyses of moderately metal-poor stars.

\item in the 3905 \AA\ line, which is very important for determining
the silicon abundances of extremely metal-poor stars, the NLTE
correction is relatively small ($\sim0.1$ dex) for moderately
metal-poor dwarfs; it increases for extremely metal-poor warm stars
($>0.15$ dex). Therefore, this line should no longer be analyzed
with LTE.

\item the 4102 \AA\ line is also important in determining the silicon
abundances of metal-poor stars; it is necessary to take account of
the NLTE effects, except for very cool and metal-rich stars. For
warm metal-poor stars, the NLTE correction can reach more than
0.25\,dex.

\item for the two \ion{Si}{ii} lines, NLTE abundance
corrections are negative. A similar behaviour is found for
\ion{Ca}{ii} lines (Mashonkina et al. \cite{MKP07}).
\end{itemize}

\begin{figure}
\resizebox{\hsize}{6.6cm}{\includegraphics[width=9.3cm]{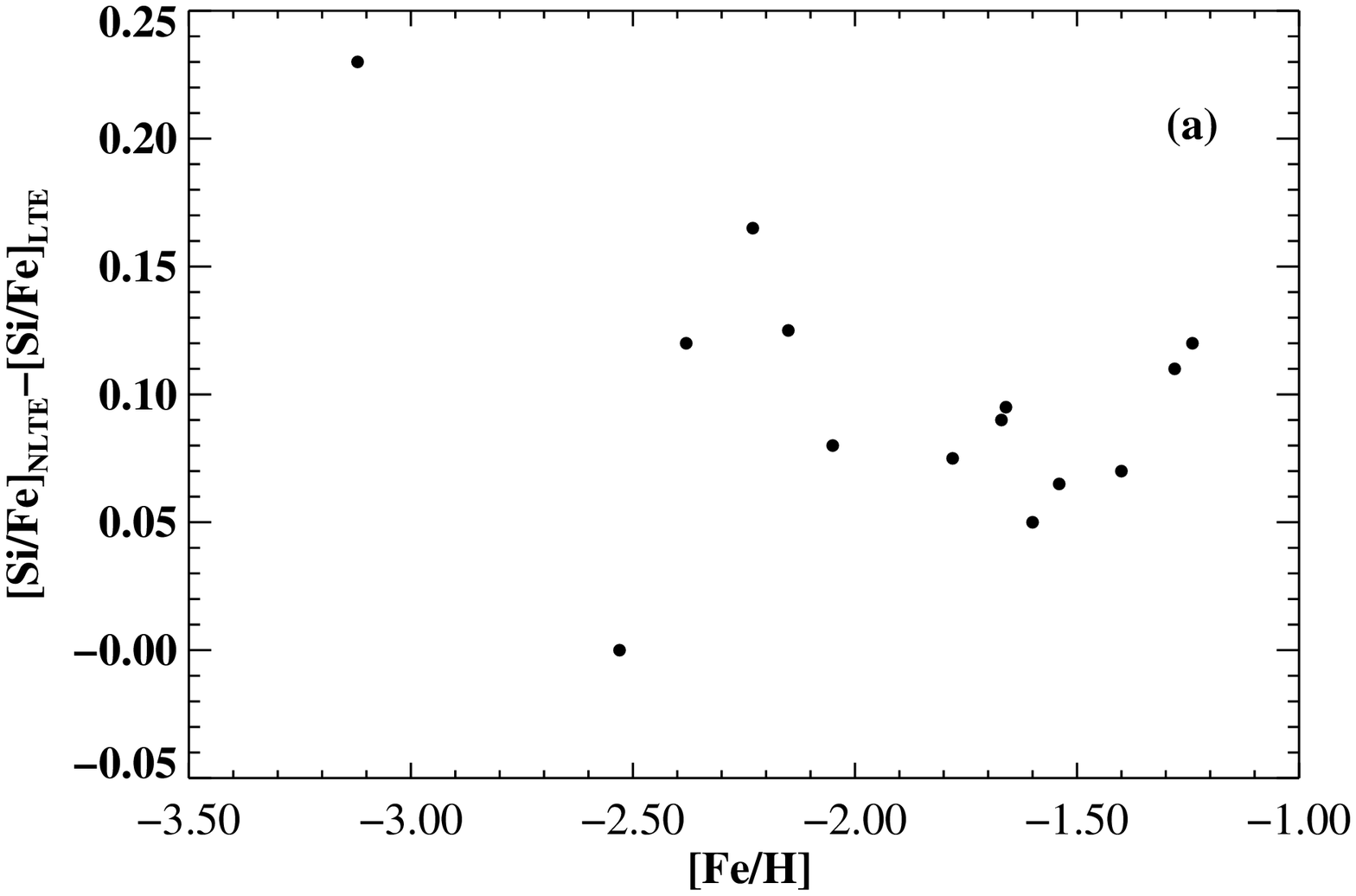}}
\resizebox{\hsize}{6.6cm}{\includegraphics[width=9.3cm]{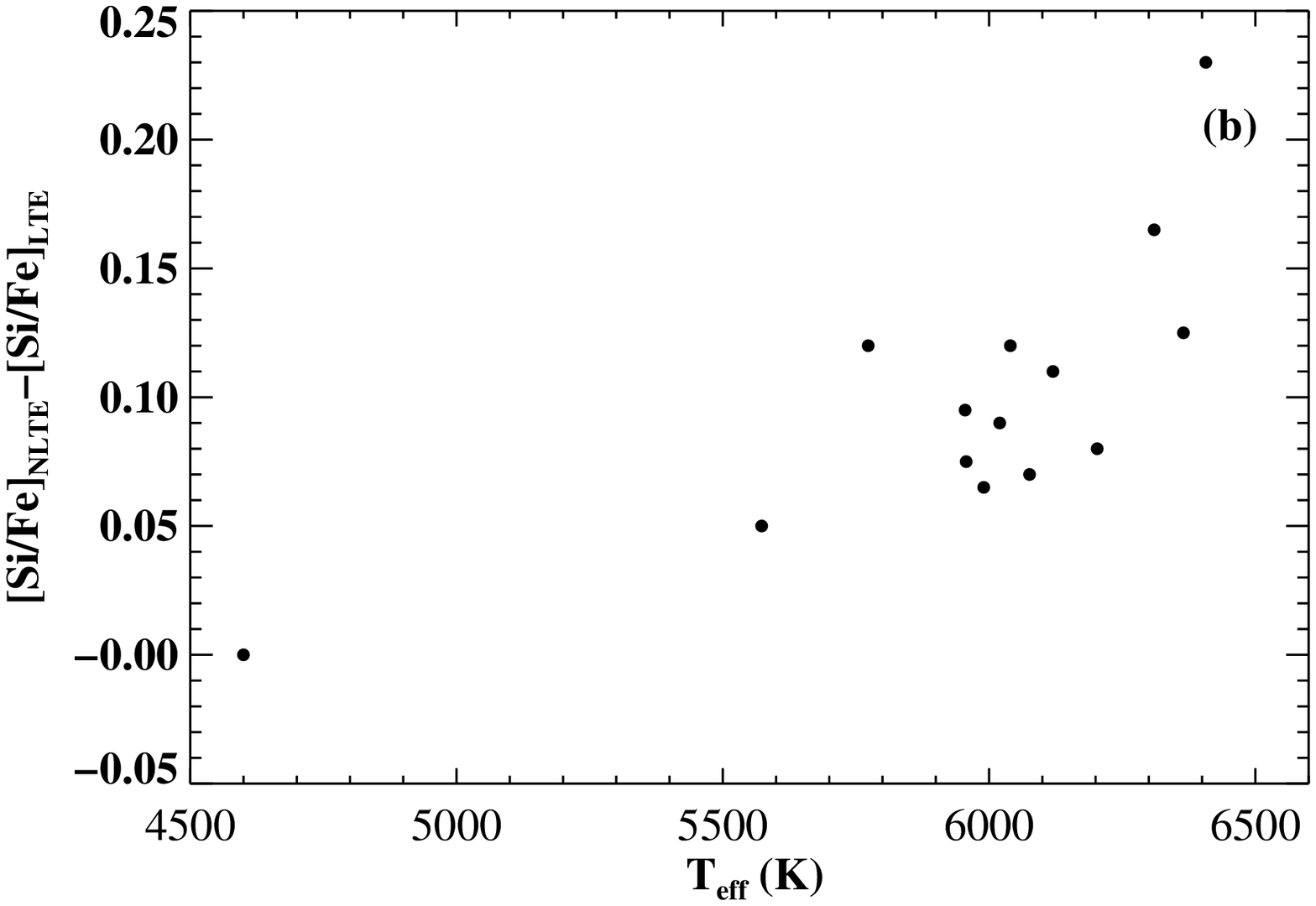}}
\resizebox{\hsize}{6.6cm}{\includegraphics[width=9.3cm]{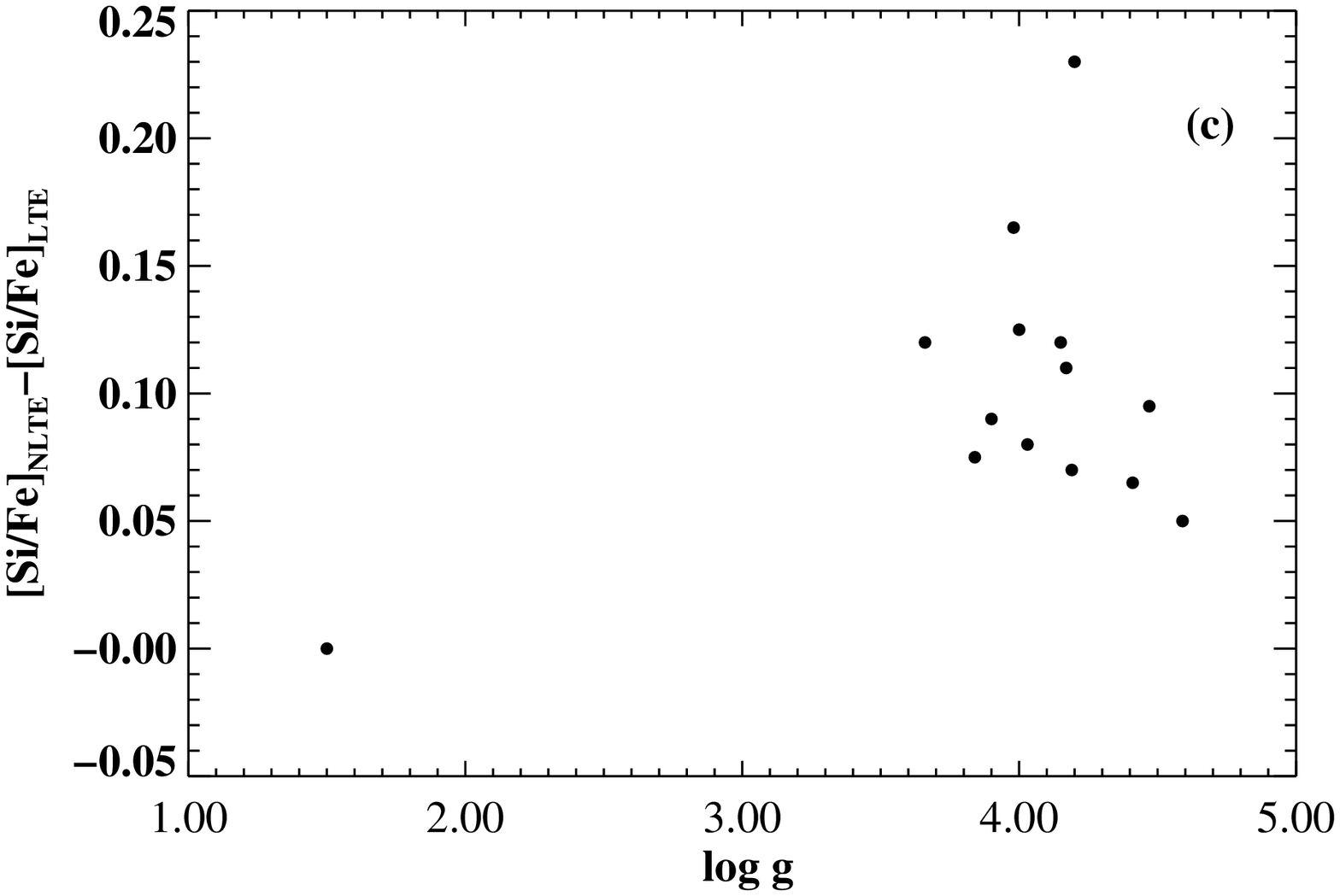}}
\caption[short title]{Difference of [Si/Fe] abundance ratios
calculated under NLTE and LTE assumptions for the two strong
\ion{Si}{i} 3905 and 4102 \AA\ lines as a function of metal
abundance (a), temperature (b), and surface gravity (c).}
\vspace*{-0.3cm} \label{fig2}
\end{figure}

Our results show that departures from LTE of the silicon level
populations appear to be larger for warm metal-poor stars. Also,
there is a clear trend that the NLTE effects increase with
increasing temperature, which can partly explain the observed
anomalous $T_{\rm eff}$ dependence of [Si/Fe] (Preston et al.
\cite{PST06}; Lai et al. \cite{LBJ08}).

\subsection{The influence of the quasi-molecular absorption in Ly$\alpha$}

Recent calculations using the R-matrix code provide detailed
photoionisation cross-sections with autoionization resonances for
the most important energy levels of \ion{Si}{i} and \ion{Si}{ii}.
For \ion{Si}{i}, the cross-sections of the lowest three terms,
\Si{3p}{3}{P}{}{}, \Si{3p}{1}{D}{}{} and \Si{3p}{1}{S}{}{} are an
order of magnitude greater than those of the next terms. It tends
to efficiently decouple the metastable terms from the excited
ones. The thresholds of the three terms are 1500, 1700 and 2000
\AA\ (Paper I). In this wavelength region, the most important
continuous absorption is the free-free quasi-molecular absorption
and satellites in Ly$\alpha$ due to collosions with H and H$^+$
(Allard et al. \cite{ADG98}), especially for warm metal-poor
stars. In Fig.\ref{fig3}, we compare the synthetic ultraviolet
(UV) fluxes for HD\,84937 without and with this process
considered. When this effect is included, the flux near 1700 \AA\
decreases by more than an order of magnitude, and it allows us to
reproduce the most characteristic UV features of warm metal-poor
stars, as illustrated by a comparison with IUE data (Cacciari
\cite{C85}) for the hot metal-poor star HD\,84937
(Fig.\ref{fig4}).

\begin{figure}
\resizebox{\hsize}{6.6cm}{\includegraphics[width=9.3cm]{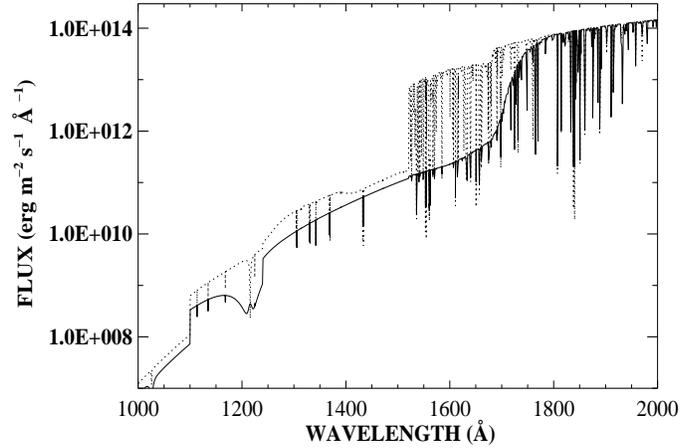}}
\caption[short title]{The UV flux emerging from MAFAGS-ODF model
for HD\,84937 without (dashed line) and with (solid line) the
Ly$\alpha$ continuum absorption.} \label{fig3}
\end{figure}

\begin{figure}
\resizebox{\hsize}{6.6cm}{\includegraphics[width=9.3cm]{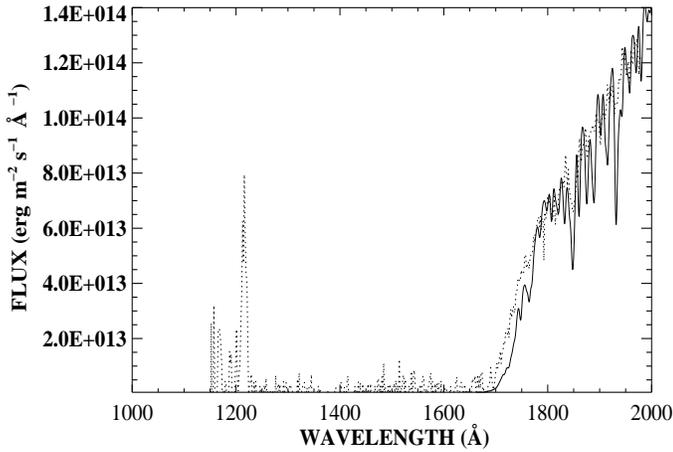}}
\caption[short title]{The comparision of the UV flux from
MAFAGS-ODF model (solid line) with the IUE observations (dashed
line) for HD\,84937.} \label{fig4}
\end{figure}
As already discussed in paper I, there is an unusually large
energy gap between the ground state, \Si{3p}{3}{P}{}{}, the two
metastable terms \Si{3p}{1}{D}{}{} and \Si{3p}{1}{S}{}{}, and the
first excited levels of neutral silicon. These gaps are $\sim$ 5,
4, and 3 eV, respectively, and they shift all lines emerging from
those levels into the UV, most of them below 2000 \AA, Therefore,
the interaction of the three most populated \ion{Si}{i} levels
with the \ion{Si}{ii} ion is completely based on bound-free
processes, where photoionisation and ionisation by electron
collisions compete in strength. Due to the large photoinisation
cross-sections for the three lowest levels, the departure
coefficients of \ion{Si}{i} are very sensitive to the UV radiation
field. When the Ly$\alpha$ continuum absorption is not included,
the extreme underpopulation of \ion{Si}{i} levels in the
atmospheres of warm metal-poor stars is clearly seen. The
depopulation with respect to the LTE case already starts in very
deep photospheric layers, with a very large net pumping to the
\ion{Si}{ii} ground state due to the strong UV radiation. For
S$_{\rm H} = 0.1$ the departure coefficients $b_i$ drop near to 0
at log $\tau$ = 0.5 (Fig.\ref{fig5}, top). As expected, the
\Si{3p}{1}{D}{}{} and \Si{3p}{1}{S}{}{} terms follow
\Si{3p}{3}{P}{}{} tightly. This would not affect neutral Si line
formation very much, except that the strong 4102 \AA\ line turns
into emission for warm metal-poor stars, such as HD\,84937, as
shown in Fig.\ref{fig6}. Also, the NLTE correction would be very
large ($\sim$ 1.0 dex) for the 3905 \AA\ line.

\begin{figure}
\resizebox{\hsize}{6.6cm}{\includegraphics[width=9.3cm]{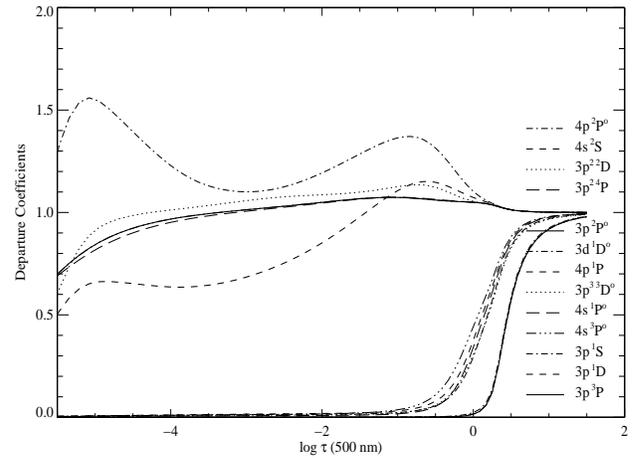}}
\resizebox{\hsize}{6.6cm}{\includegraphics[width=9.3cm]{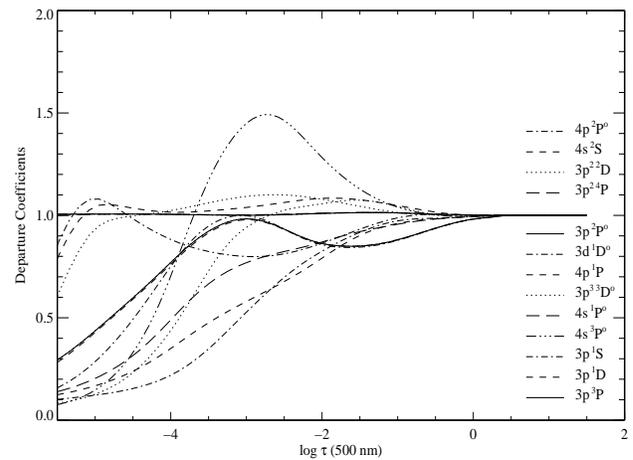}}
\caption[short title]{Departure coefficients as a function of log
$\tau$ for selected levels of \ion{Si}{i} and \ion{Si}{ii} for
HD\,84937 without (top) and with (bottom) the Ly$\alpha$ continuum
absorption.} \label{fig5}
\end{figure}

Models with Ly$\alpha$ continuum absorption included produce much
smaller departures (Fig.\ref{fig5}, bottom), as radiative
processes (photoionisation) are compensated by hydrogen and
electron collisions. Since all the excited terms are only loosely
coupled to each other and to other high-excitation terms, their
departure coefficients tend to diverge at $\log\tau \simeq -2$.
More importantly, similar abundances can be obtained from both the
3905 and 4102 \AA\ lines.

\begin{figure}
\resizebox{\hsize}{6.6cm}{\includegraphics[width=9.3cm]{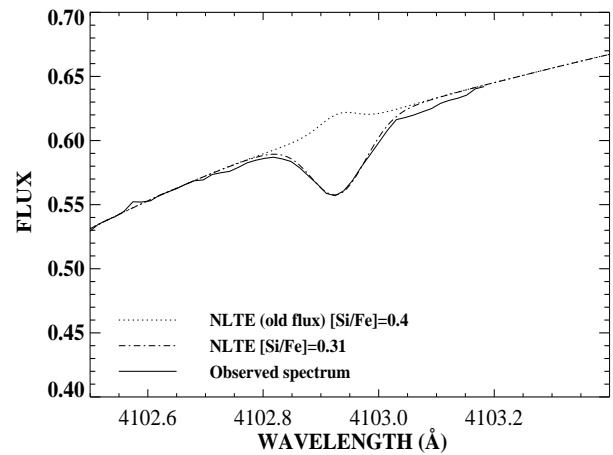}}
\caption[short title]{Theoretical NLTE flux profiles (\SH\ =0.1)
without (dotted lines) and with (dash dotted lines) the Ly$\alpha$
continuum absorption.} \label{fig6}
\end{figure}
\section{Results}

\subsection{Stellar silicon abundances}
The abundance determinations for our program stars are made using
spectral synthesis. The synthetic spectra are convolved with
macroturbulence, rotational and instrumental broadening profiles, in
order to match the observed spectral lines. Considering the NLTE
abundances, our results do not show a large abundance discrepancy
between different lines; the final abundance scatter of single lines
is between 0.01 and 0.14. The derived abundances are presented in
Table \ref{table2}.

\subsection{Comparison with other work}
Silicon abundances for metal-poor stars have been determined by
several groups based on LTE analyses. In Fig.\ref{fig7}, we
compare the [Si/Fe] values (NLTE) determined in this paper with
those from the literature. Some systematic differences can be seen
from this figure. In the remaining part of this section we will
briefly discuss some possible reasons for these differences.

\begin{figure}
\resizebox{\hsize}{6.6cm}{\includegraphics[width=9.3cm]{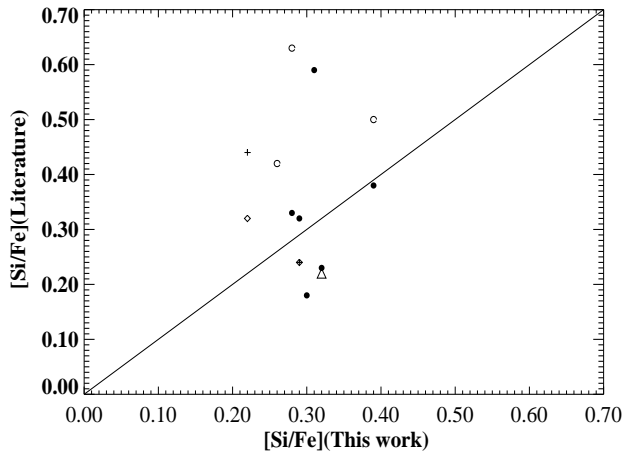}}
\caption[short title]{Comparison of derived [Si/Fe] (NLTE) for
stars in common with other studies. Triangle ($\triangle$) is from
Decauwer et al. (\cite{DJP05}); open circles ($\circ$) are from
Fulbright (\cite{Fu00}); Filled circles ($\bullet$) are from
Gratton et al. (\cite{GCC03}); Pluses ($+$) are from Honda et al.
(\cite{H04}); Diamonds ($\diamond$) are Ryan et al.
(\cite{RNB96})} \vspace*{-0.3cm} \label{fig7}
\end{figure}

\vskip 0.2cm
 \noindent{\underline{Decauwer et al. (\cite{DJP05})}}
\vskip 0.1cm

Decauwer et al. (\cite{DJP05}) performed a LTE line formation
analysis for a sample of moderately metal-poor stars, and they used
weak \ion{Si}{i} lines when determining the Si abundances. We have
one star in common with theirs. Our [Si/Fe] is about 0.1 dex higher
than theirs. The $gf$ values adopted in their study are nearly the
same as those used here. We note that the optical \ion{Si}{i} lines
are very weak in metal-poor stars, so the determination of the
continuum is difficult, while the NLTE effects for these lines can
be negligible. The difference is mostly due to the determination of
the continuum.

\vskip 0.2cm \noindent{\underline{Fulbright (\cite{Fu00},
\cite{Fu02})}} \vskip 0.1cm

This analysis deals with a large number of metal-poor stars, of
which three stars are in common with our sample. For the three stars
in common with our list we obtain $\overline{\Delta
\rm[Si/Fe]}$=$-0.21\pm0.13$. We find that the greatest difference
($\sim$0.35 dex) comes from metal-poor stars such as HD\,31128. They
also used weak \ion{Si}{i} lines in determining the silicon
abundances, and took the log $gf$ values from the laboratory work of
Garz (\cite{Garz73}). Their $gf$ values are about 0.1 dex lower then
ours. Also, their effective temperatures are about 200 K lower.
Thus, the difference in $gf$ values and temperatures can explain the
discrepancies.

\vskip 0.2cm \noindent{\underline{Gratton et al.(\cite{GCC03})}}
\vskip 0.1cm The authors determined the silicon abundances for 150
field subdwarfs and subgiants. Our results are mostly in agreement
with theirs. For the six stars in common, the average difference is
$0.003\pm0.09$.

\vskip 0.2cm
 \noindent{\underline{Honda et al. (\cite{H04})}}
  \vskip 0.1cm

Using high resolution, high signal to noise ratio spectra of 22 very
metal-poor stars from Subaru/HDS, the authors confirmed the
overabundance ratio [Si/Fe] of metal-poor giants found previously by
Cayrel et al. (\cite{CDS04}). Their results were determined from the
4102 \AA\ line; the $gf$ value adopted in this study is also from
the laboratory work of Garz (\cite{Garz73}). It is 0.15\,dex lower
than ours. We have two stars in common with this work; the giant
star HD\,122563 and the subgiant star HD\,140283. For the giant, the
silicon abundance determined by this study is 0.22 dex higher than
ours; most of the difference comes from the different $gf$ value
adopted. For the subgiant, their result is -0.04\,dex lower than
ours; this is due to the large NLTE effects ($+$0.19\,dex) in this
star. Thus, the difference in $gf$ value is compensated for by the
large NLTE effect.

\vskip 0.2cm \noindent {\underline{Ryan et al. (\cite{RNB96})}}
\vskip 0.1cm

This work analysed 19 metal-poor stars using the 3905 \AA\ line. The
$gf$ value adopted in this study was also from the laboratory work
of Garz (\cite{Garz73}), and it is nearly same as the value we
adopted. The results of Ryan et al. (\cite{RNB96}) are in agreement
with ours. For the two common stars from Ryan et al. (\cite{RNB96}),
the average difference between theirs and ours is $0.025\pm 0.04$.
Part of the difference is due to the NLTE effects for this strong
line.

\section {Discussion}

\subsection{\ion{Si}{i}/\ion{Si}{ii} ionization equilibrium in metal-poor
stars}

There are nine stars in our sample with Si abundances determined
from both \ion{Si}{i} and \ion{Si}{ii} lines, and excellent
agreement between the two ionization stages is achieved. The average
difference of $\Delta$[\ion{Si}{i}/Fe]-[\ion{Si}{ii}/Fe] is $0.001
\pm 0.04$. The largest difference between \ion{Si}{i} and
\ion{Si}{ii} abundances is found to be 0.09 for CD\,$-$51$^\circ$
4628. It may due to the large uncertainty in the HIPPARCOS parallax
for this star ($\sim$ 20\%).
%Can the stellar parameters cause this discrepancy?
We conclude that, within the modelling uncertainties, NLTE leads to
consistent Si abundances derived from the two ionization stages,
while LTE fails to give consistent results.
%We show that NLTE
%largely removes obvious discrepancies between \ion{Si}{i} and
%\ion{Si}{ii}.

\subsection{Silicon abundance and nucleosynthesis in the early Galaxy}

The variation of [Si/Fe] with the stellar metallicity [Fe/H]
contains information about the chemical evolution of the Galaxy.
Fig. \ref{fig8} displays the behaviour of the [Si/Fe] ratio
(calculated in LTE and NLTE) with the metal abundance for all
stars considered in this paper. One important feature that can be
seen from Fig. \ref{fig8} is that there is a large scatter in the
LTE results, while when the NLTE effects are included, the bulk of
[Si/Fe] ratio is $\sim$ 0.3 for our program stars. The higher
[Si/Fe] ratio at lower metallicity found in some studies of
metal-poor stars (e.g. Fulbright \cite{Fu02}, Stephens \&
Boesgaard \cite{SB02}) is not confirmed by this work. The obvious
exceptions are the two moderately metal-poor stars
(CD\,$-$51$^\circ$ 4628 and HD\,122196), which exhibit relatively
low [Si/Fe] values. We note that these two stars also show lower
values of magnesium. Such stars were also found by Nissen \&
Schuster (\cite{NS97}) and Cohen et al. (\cite{C07}, \cite{C08}).

Detailed modelling of Galactic chemical evolution has been attempted
by many authors (e.g. Timmes et al. \cite{TWW95}; Goswami \&
Prantzos \cite{GP00}; Kobayashi et al. \cite{KUN06}). Based on
Woosley \& Weaver's (\cite{WW95}) metallicity-dependent yields,
Timmes et al. (\cite{TWW95}) calculate the behaviour of [Si/Fe] as a
function of metallicity. Their result predicts that [Si/Fe]
increases from [Fe/H] $\sim$ $-0.5$ to $\sim -1.5$, while it
decreases from [Fe/H] $\sim$ $-2$ to $\sim -3$ (see their Fig. 21).
The decrease of [Si/Fe] with decreasing metallicity is due to
(unspecified) uncertainties in the low-metallicity massive star
models. However, using the same yields but with an iron yield
reduced by a factor of 2, Goswami \& Prantzos (\cite{GP00}) show
that the [Si/Fe] increases from [Fe/H]$\sim$ $0$ to [Fe/H] $\sim
-4$. They predict that the [Si/Fe] ratio is about 0.5 dex at
[Fe/H]$=-$3. Their result suggests an increasing [Si/Fe] towards the
lower metallicity regime. They used a different initial mass
function and a different halo model. A similar result was found by
Kobayashi et al. (\cite{KUN06}, see their Fig.10). The Si yields of
this work are larger for more massive metal-free stars.

Using magnesium instead of iron as the reference can remove Type Ia
SNe from consideration, because most of the silicon and nearly all
the magnesium are produced in massive stars (Timmes et al.
\cite{TWW95}). Calculations of nucleosynthesis for SNe Ia show that
they should produce some silicon (Nomoto et al. \cite{NIN97}). The
overall behaviour of the [Si/Mg] ratios versus [Fe/H] is shown in
Fig. \ref{fig9}, where magnesium abundances are taken from Gehren et
al. (\cite{GLS04}, \cite{GSZ06}) and Mashonkina et al.
(\cite{MZG08}). Although there is a scatter, our result shows that
the [Si/Mg] ratio is around $-$0.1.

Our observational results provide some implications for the
nucleosynthesis of silicon. Theoretically, it is expected that
silicon is synthesized predominantly in moderate mass ($\sim$ 20 $M
_{\sun}$) Type II SNe, while magnesium is produced primarily in the
higher mass ($\sim$ 35 $M_{\sun}$) Type II SNe (Woosley \& Weaver
\cite{WW95}). The overabundance of silicon and the ratio of [Si/Mg]
is about $-0.1$ in metal-poor stars, suggests that, similar to
magnesium, silicon is produced by massive type II SNe. The slightly
lower [Si/Fe] values for extremely metal-poor stars may indicate
that silicon is produced more by moderately massive type II SNe.
This result agrees with nucleosynthesis calculations of massive
stars by Woosley \& Weaver (\cite{WW95}), but it is at variance with
the calculations by Nomoto et al. (\cite{NTU06}). It should be noted
that there is only one extremely metal-poor star in our sample; it
is necessary to investigate more extreme metal-poor stars to confirm
this suggestion. From a theoretical point of view, since silicon is
produced in a deeper region than magnesium, the ratio of Mg/Si is
sensitive to the explosion energy and the outer border of the mixed
region. The higher explosion energy yields a smaller Mg/Si ratio,
while the Mg/Si ratio decreases with the increasing outer boundary
of the mixed region (Umeda \& Nomoto \cite{UN05}).

\begin{figure}
\resizebox{\hsize}{6.6cm}{\includegraphics[width=9.3cm]{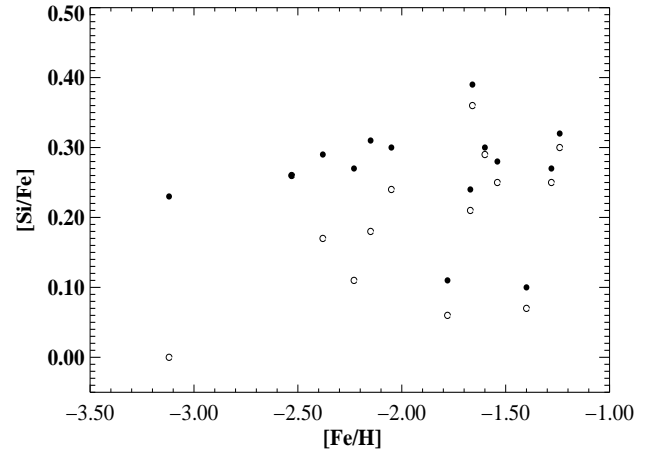}}
\caption[short title]{Abundance ratios [Si/Fe] as a function of
[Fe/H]. Filled circles ($\bullet$) represent NLTE analysis, while
open circles ($\circ$) for LTE analysis.} \label{fig8}
\end{figure}

\begin{figure}
\resizebox{\hsize}{6.6cm}{\includegraphics[width=9.3cm]{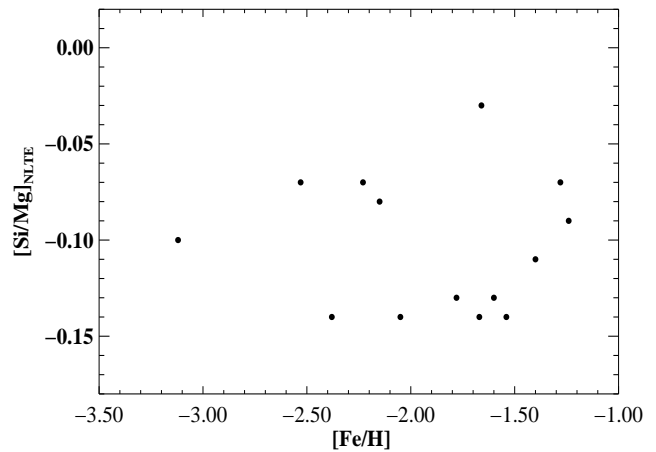}}
\caption[short title]{Abundance ratios [Si/Mg] for NLTE analysis
as a function of [Fe/H].} \label{fig9}
\end{figure}

\subsection{Comparison with DLA abundances}

%Damped Ly$\alpha$ (DLA) absorption systems seen in the spectra of
%QSOs are presumally the result of absorption from a background QSOs.
The chemistry of high-redshift DLA systems should represent at least
crudely the state of the ISM gas at an early stage in the formation
of the halo of the Galaxy. It is expected that the abundance ratios
of elements determined from DLA would be consistent with those
deduced from spectra of extremely metal-poor stars in our Galacy,
although the metallicities of DLA systems do not reach values as low
as those of individual Galactic halo stars. The Si/Fe ratio is of
crucial importance in understanding the chemical nature of DLAs
(Henry \& Prochaska \cite{HP07}).

As DLA systems are generally in young evolutionary stages, it is
expected that supersolar [Si/Fe] values will be observed. Prochaska
\& Wolfe (\cite{PW02}) found that [Si/Fe] values exhibt a plateau
$\simeq 0.3$ dex at [Si/H] $<-1.5$, and Dessauges-Zavadsky et al.
(\cite{DPD06}) found an average [Si/Fe] $\simeq 0.43$ dex for their
sample, which indicates significant $\alpha$-enrichment in the DLA
systems at low metallicity. Based on a large number of high
precision measurements obtained with \'{e}chelle spectrometers on 8-
to 10-m-class telescopes, Henry \& Prochaska (\cite{HP07}) also
found that [Si/Fe] for all their objects is $+$0.44. However, the
mean [Si/Fe] predicted by their model for their sample is $+$0.31
dex, and the ratio is roughly constant. This is very similar to our
observational results, but is slightly lower than the value obtained
from DLA systems. The most likely explanation is Fe depletion onto
dust (Savage \& Sembach \cite{SS96}) in DLA systems.

\section{Conclusions}
We have determined silicon abundances for 14 metal-poor stars,
spanning the range -3.5 $<$ [Fe/H] $<$-1.4. All abundances are
derived from NLTE statistical equilibrium calculations. Based on our
results we come to the following conclusions:
\begin{enumerate}
\item The [Si/Fe] ratios are overabundant for metal-poor stars, and
there is an indication that [Si/Fe] slightly decreases with
decreasing metallicity.

\item The NLTE effects are different from line to line. The weak lines are
insensitive to NLTE effects, while the strong 3905 and 4102 \AA\
lines show large NLTE effects. Large departures from LTE appear in
warm metal-poor stars. The NLTE effects increase with increasing
temperature, which can partly explain the observed results of an
anomalous $T_{\rm eff}$ dependence of [Si/Fe] (Preston et al. 2006;
Lai et al. 2008).

\item A different situation is found for \ion{Si}{ii}. For the two \ion{Si}{ii}
lines, NLTE leads to enhanced absorption in the line cores and
negative abundance corrections over the range of stellar parameters
studied here.

\item The Ly$\alpha$ continuum absorption and the radiative bound-free
cross-sections are very important for Si line formation, especially
for warm metal-poor stars.

\item Taking advantage of our statistical equilibrium approach and accurate
atomic data for the investigated lines, we obtain good agreement
between the \ion{Si}{i} and \ion{Si}{ii} abundances. We show that
NLTE largely removes obvious discrepancies between \ion{Si}{i} and
\ion{Si}{ii} obtained under an LTE assumption.

\item  Our results suggest that, similar to magnesium, silicon is produced by massive
type II SNe.
\end{enumerate}

It would be important to perform NLTE abundance determinations for
silicon for some warm extremely metal-poor stars.

\begin{acknowledgements}
J.R. acknowledges Dr. Aoki for providing the Subaru spectrum for
G64-12. This research was supported by the National Natural Science
Foundation of China under grant Nos. 10821061, 10778626, 10811130393
and the National Basic Research Program of China (973 Program) under
grant No. 2007CB815103, by the Deutsche Forschungsgemeinschaft with
grant 446 CHV 112/1,2 and the Russian Foundation for Basic Research
with grant 08-02-92203-NNSF.
\end{acknowledgements}

% end of the main text
\clearpage
\onecolumn
%\begin{onecolumn}
\begin{table}
\scriptsize \caption[2]{
%Atmospheric parameters and \textbf{[Si/Fe]}
%of the program stars. For each star the first row indicates the LTE
%results, while the second row indicates the NLTE results. Stellar
%parameters are the same as in Gehren et al. (\cite{GLS04},
%\cite{GSZ06}) and Mashokina et al. (\cite{MZG08}). \textbf{The iron
%abundances are derived from \ion{Fe}{ii} lines.
Stellar silicon LTE and NLTE (for each star, the first and the
second row, respectively) abundances given relative to the iron
LTE abundances derived from the Fe II lines$^*$.}
\setlength{\tabcolsep}{0.1cm}
\begin{tabular}{lccccrrrrrrrrrrrrrll}
\hline\hline\noalign{\smallskip}
   Name   & $T {\rm _{eff}}$  & $\log g$ & [Fe/H] &$\xi$   &   3905 & 4102  & 5690 & 5701 & 5772&  6142&  6145&  6155&
   6237&  6243 & 6244 &6347 & 6371 &[\ion{Si}{i}/Fe] &[\ion{Si}{ii}/Fe]\\
\noalign{\smallskip}\hline\noalign{\smallskip}
BD\,$-$04$^\circ$ 3208&  6310&  3.98 & -2.23 & 1.50 &   0.22 &-0.01  &      &      &     &      &      &      &      &       &      &     &      & 0.11$\pm$0.12& \\
          &      &      &       &      &   0.30 & 0.24  &      &      &     &      &      &      &      &       &      &     &      & 0.27$\pm$0.03&\\
CD\,$-$51$^\circ$ 4628&  6076& 4.19 & -1.40 & 1.50 &   0.08 &-0.03  &      & 0.09 &     &      &      &  0.13&      &    &  & 0.19&  0.18& 0.07$\pm$0.05& 0.19$\pm$0.01\\
          &      &      &       &      &   0.10 & 0.09  &      & 0.09 &     &      &      &  0.13&      &       &      & 0.19&  0.18& 0.10$\pm$0.01& 0.19$\pm$0.01\\
HD\,29907   &  5573& 4.59 & -1.60 & 0.90 &   0.31 & 0.23  &      & 0.32 & 0.28&      &      &  0.29&  0.31&  0.28 & 0.28 & 0.31&      & 0.29$\pm$0.02& 0.31\\
          &      &      &       &      &   0.35 & 0.29  &      & 0.32 & 0.28&      &      &  0.29&  0.31&  0.28 & 0.28 & 0.30&      & 0.30$\pm$0.02& 0.30\\
HD\,31128   &  5990& 4.41 & -1.54 & 1.30 &   0.34 & 0.22  &      &      &     &      &      &  0.24&  0.24&  0.24 & 0.24 & 0.24&      & 0.25$\pm$0.03& 0.24\\
          &      &      &       &      &   0.35 & 0.34  &      &      &     &      &      &  0.24&  0.24&  0.24 & 0.24 & 0.23&      & 0.28$\pm$0.03& 0.23\\
HD\,34328   &  5955& 4.47 & -1.66 & 1.30 &   0.37 & 0.26  & 0.40 &      &     &      &      &  0.34&  0.37&  0.41 & 0.39 & 0.38&  0.39& 0.36$\pm$0.04& 0.39$\pm$0.01\\
          &      &      &       &      &   0.42 & 0.40  & 0.40 &      &     &      &      &  0.34&  0.37&  0.41 & 0.39 & 0.36&  0.37& 0.39$\pm$0.02& 0.37$\pm$0.01\\
HD\,59392   &  6020& 3.90 & -1.67 & 1.70 &   0.24 & 0.08  &      &      &     &      &      &  0.21&  0.22&  0.24 & 0.24 & 0.26&  0.24& 0.21$\pm$0.04& 0.25$\pm$0.01\\
          &      &      &       &      &   0.26 & 0.24  &      &      &     &      &      &  0.21&  0.22&  0.24 & 0.24 & 0.24&  0.23& 0.24$\pm$0.01& 0.24$\pm$0.01\\
HD\,74000   &  6203& 4.03 & -2.05 & 1.70 &   0.26 & 0.19  &      &      &     &      &      &  0.28&      &       &      & 0.30&      & 0.24$\pm$0.04& 0.30\\
          &      &      &       &      &   0.30 & 0.31  &      &      &     &      &      &  0.28&      &       &      & 0.28&      & 0.30$\pm$0.01& 0.28\\
HD\,84937   &  6365& 4.00 & -2.15 & 1.60 &   0.25 & 0.11  &      &      &     &      &      &      &      &       &      &     &      & 0.18$\pm$0.07&\\
          &      &      &       &      &   0.30 & 0.31  &      &      &     &      &      &      &      &       &      &     &      & 0.31$\pm$0.01&\\
HD\,97320   &  6040& 4.15 & -1.24 & 1.30 &   0.32 & 0.15  & 0.35 & 0.32 & 0.31&  0.34&  0.33&  0.27&  0.31&  0.31 & 0.31 & 0.31&  0.34& 0.30$\pm$0.03& 0.33$\pm$0.02\\
          &      &      &       &      &   0.37 & 0.34  & 0.35 & 0.32 & 0.31&  0.34&  0.33&  0.27&  0.31&  0.31 & 0.31 & 0.30&  0.34& 0.32$\pm$0.02&
          0.32$\pm$0.02\\
HD\,102200  &  6120& 4.17 & -1.28 & 1.50 &   0.30 & 0.10  & 0.34 & 0.25 & 0.26&  0.27&  0.28&  0.20&  0.20&  0.31 & 0.29 & 0.25&  0.30& 0.25$\pm$0.05& 0.28$\pm$0.03\\
          &      &      &       &      &   0.34 & 0.28  & 0.34 & 0.25 & 0.26&  0.27&  0.28&  0.20&  0.20&  0.31 & 0.29 & 0.23&  0.28& 0.27$\pm0.04$& 0.28$\pm$0.01\\
HD\,122196  &  5957& 3.84 & -1.78 & 1.70 &   0.12 &-0.04  &      &      &     &      &      &  0.09&      &       &      & 0.13&  0.13& 0.06$\pm$0.06& 0.13\\
          &      &      &       &      &   0.13 & 0.10  &      &      &     &      &      &  0.09&      &       &      & 0.11&  0.11& 0.11$\pm$0.02& 0.11\\
HD\,122563  &  4600& 1.50 & -2.53 & 1.90 &   0.24 & 0.19  &      &      &     &      &      &      &      &       &      &     &      & 0.22$\pm$0.05&\\
          &      &      &       &      &   0.24 & 0.19  &      &      &     &      &      &      &      &       &      &     &      & 0.22$\pm$0.05&\\
HD\,140283  &  5773& 3.66 & -2.38 & 1.50 &   0.27 & 0.07  &      &      &     &      &      &      &      &       &      &     &      & 0.17$\pm$0.10&\\
          &      &      &       &      &   0.32 & 0.26  &      &      &     &      &      &      &      &       &      &     &      & 0.29$\pm$0.03&\\
G\,64-12    &  6407& 4.20 & -3.12 & 2.30 &  -0.02 &       &      &      &     &      &      &      &      &       &      &     &      & -0.02&\\
          &      &      &       &      &   0.23 &       &      &      &     &      &      &      &      &       &      &     &      & 0.23&\\
\noalign{\smallskip} \hline
\end{tabular}

$^*$ Our NLTE calculations for Fe I/II based on the advanced
atomic model (Mashonkina et al. \cite{MGS09}) support the earlier
conclusion of Korn et al. (\cite{KST03}) that the NLTE effects for
the Fe II lines are negligible.
 \label{table2}
\end{table}
%\end{onecolumn}

\end{document}